\documentclass[12pt]{iopart}
\usepackage{iopams}
\usepackage{url}
\usepackage{graphicx,color}
\usepackage{bbm}

\expandafter\let\csname equation*\endcsname\relax
\expandafter\let\csname endequation*\endcsname\relax
\usepackage{amsmath}

\newcommand{\jk}[1]{\textcolor{black}{{#1}}}

\begin{document}
\title[Epidemic compartment model with asymptomatic infections and mitigation]{Nonlinear dynamics of an epidemic compartment model with asymptomatic infections and mitigation}
\author{Maurice G\"ortz and Joachim Krug}
\address{Institute for Biological Physics, University of Cologne,
  Z\"ulpicher Strasse 77, 50937 K\"oln, Germany}
\vspace{10pt}
\begin{indented}
\item[]\today
\end{indented}
\begin{abstract}
  A significant proportion of the infections driving the current {SARS-CoV-2} pandemic are transmitted asymptomatically.
  Here we introduce and study a simple epidemic model with separate compartments comprising asymptomatic and symptomatic
  infected individuals. The linear dynamics determining the outbreak condition of the model is equivalent to
  a renewal theory approach with exponential waiting time distributions. Exploiting a nontrivial conservation law of the
  full nonlinear dynamics, we derive analytic bounds on the peak number of infections in the absence and presence of mitigation
  through isolation and testing. The bounds are compared to numerical solutions of the differential equations. 
\end{abstract}

\submitto{\jpa}

\section{\label{Sec:intro}Introduction}

The ongoing SARS-CoV-2 pandemic highlights the importance of presymptomatic and asymptomatic infections
\cite{Rothe2020,Huang2021,Byambasuren2020,Alene2021,Oran2021,Meyerowitz2021,Dobrovolny2020,Koelle2022} and their effects on the ability to control outbreaks by non-pharmaceutical interventions \cite{Ferretti2020,Hart2021,Contreras2021,Tian2021}. These effects are principally twofold. On the one hand, asymptomatic infections often remain undetected \cite{Li2020}, which makes it difficult to monitor the spread of the disease in the population and confounds estimates of epidemiological parameters \cite{Park2020}. On the other hand, asymptomatic individuals do not require medical treatment and therefore do not contribute to the burden on the health care system caused by the epidemic.

A mathematical framework for modeling epidemics with asymptomatic infections based on renewal theory was developed by
Fraser et al. in the context of the first SARS pandemic \cite{Fraser2004}. In this approach, the transmission
of the disease is described by two functions $\beta(\tau)$ and $\sigma(\tau)$ that
quantify the infectiousness of an indvidual and the probability for
the individual to still be asymptomatic,
respectively, as a function of the time $\tau$ since infection. Both
functions can be estimated from empirical data
\cite{Ferretti2020,Tian2021}. For a given set of functions $\beta, \sigma$ the theory provides
a condition for an outbreak to occur, and quantifies the efficacy of isolation and contact tracing measures required for
controlling it. As usual, the outbreak criterion is determined by the linear dynamics in the early stages of the epidemic.
It takes the form $R_0 > 1$, where $R_0$ denotes the basic
reproduction number defined as the expected number of secondary
  infections conferred by an infected individual in a fully
  susceptible population \cite{Murray,Hethcote2000,Keeling2008,Grassly2008}.  

However, in order to predict the severity of an outbreak, it is necessary to understand the nonlinear epidemic dynamics, which
determines key quantities such as the total number of infections at the end of the outbreak \cite{Ma2006,Leung2018} or the number
of infected individuals at the peak of the epidemic. The purpose of this contribution is to introduce and study a simple,
analytically tractable model that allows to address the effect of asymptomatic infections on the full nonlinear time evolution
of an epidemic. The model is a minimal extension of the standard SIR-model \cite{Murray,Hethcote2000,Keeling2008,Grassly2008,Traulsen2022}
that includes separate populations of asymptomatic and symptomatic infected individuals; 
similar models have been introduced previously and are sometimes referred to as SAIR models, see \cite{Ansumali2020} and 
\ref{Sec:Related}.

The specific version of the SAIR model considered in this work is defined in the next section. In
Sections \ref{Sec:Linear} and \ref{Sec:Parameters} we analyze the linear dynamics 
and identify the basic reproduction number, the fraction of asymptomatic infections and the
functions $\beta(\tau)$ and $\sigma(\tau)$.  An extended model that includes the recovery of asymptomatic
  individuals will be discussed in Sect.~\ref{Sec:AtoR}.
In Sect.~\ref{Sec:Peak} we make
use of the nontrivial conservation law of the dynamical system to derive an analytic upper bound on the peak number of symptomatically
infected individuals, which is compared to numerical solutions. Section \ref{Sec:Interventions} generalizes the model to
include the effects of isolation (of symptomatic individuals) and testing (of asymptomatic individuals), and Section
\ref{Sec:Conclusions} summarizes our conclusions.

\section{Epidemic dynamics with asymptomatic infections}

\subsection{Dynamical equations}
\label{Sec:SAIR}

\begin{figure}[htb]
\begin{center}
\includegraphics[width=0.9\textwidth]{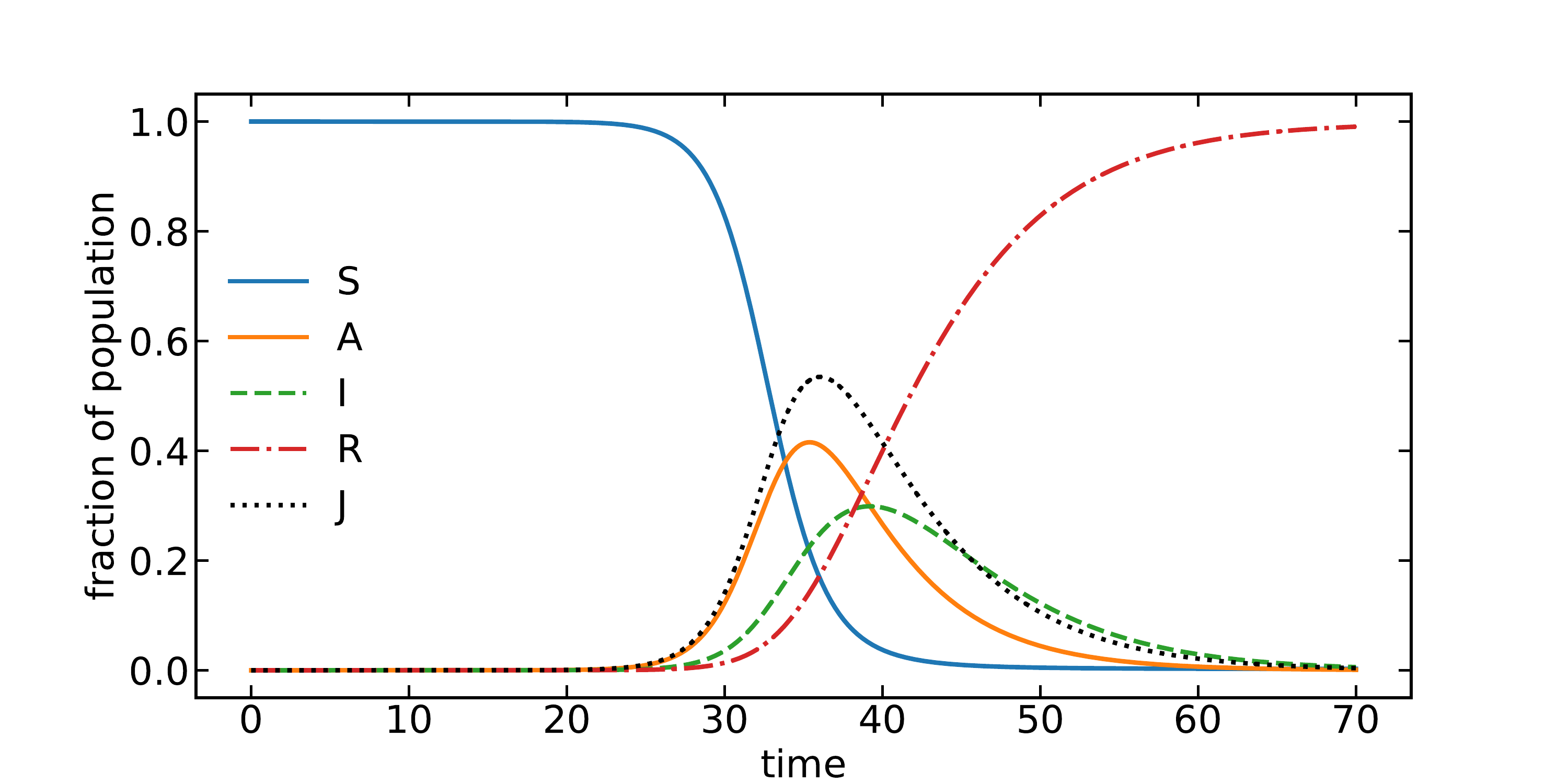}
\end{center}
\caption{\label{Fig0} Illustration of the epidemic dynamics described
  by Eqs.~(\ref{S}-\ref{R}). The figure shows the time courses of
  $S(t)$, $A(t)$, $I(t)$ and $R(t)$, as well as the auxiliary quantity $J(t)$
  defined in Eq.~(\ref{Jdef}), which peaks when $S = S^\ast = \frac{N}{R_0}$. Dimensionless parameters are $R_0 = 6$ and
  $\theta = 0.5$, and the mean infection time is $\mathbb{E}(\tau) = 10$. The initial condition consists of one
asymptomatically infected individual in a population of size $N = 10^8$.} 
\end{figure}

We consider a population of constant size $N$ which is subdivided into
compartments comprising susceptible ($S$), asymptomatically infected ($A$),
symptomatically infected ($I$) and removed ($R$) individuals. For the
sake of simplicity, asymptomatic and symptomatic individuals are taken
to infect
the susceptible individuals at the same rate $\lambda$. Moreover,
following previous work \cite{Tian2021,Fraser2004}, we assume that all asymptomatic
individuals develop symptoms and assign a rate $\xi$ to this
process. We shall see
below in Sec.~\ref{Sec:Parameters} that the probability $\sigma(\tau)$ that 
an infected individual has not developed symptoms up to time $\tau$
may nevertheless remain nonzero for $\tau \to \infty$.
Symptomatically infected individuals are transferred to the removed
compartment (by recovery or death) at rate $\mu$. Denoting the
  number of individuals in the different compartments by $S(t)$,
  $A(t)$, $I(t)$ and $R(t)$, respectively, this leads to the
following set of differential equations:  
\begin{eqnarray}
\dot{S} =  - \frac{\lambda}{N} (A+I) S  \label{S} \\
  \dot{A}  =  \frac{\lambda}{N} (A+I) S - \xi A \label {A}\\
  \dot{I}  =  \xi A - \mu I \label{I} \\
  \dot{R}  =  \mu I \label{R}
\end{eqnarray}
These equations define a special case of a larger class of models
with asymptomatic infections that are 
reviewed, e.g., in \cite{Chisholm2018} and in \ref{Sec:Related}. Here we
focus on the minimal version (\ref{S}-\ref{R}) that extends the SIR model by a single additional parameter, $\xi$, which is
related to the fraction of asymptomatic infections (see
Sect.~\ref{Sec:Parameters}). The SIR model is recovered in the limit $\xi \to
\infty$. Figure \ref{Fig0} shows an exemplary numerical solution of the system (\ref{S}-\ref{R}). 
      
\subsection{Distribution of infection times}
\label{Sec:Linear}

\jk{Let $q(\tau)$ denote the probability that an individual infected at time $t$ is still infectious at time $t + \tau$. An infectious individual
is either asymptomatic or symptomatic, and the transitions $A \to I$ and $I \to R$ occur at rates $\xi$ and $\mu$, respectively.
We can thus write $q(\tau) = q_A(\tau) + q_I(\tau)$, where the contributions $q_A$ and $q_I$ from asymptomatic and symptomatic
individuals satisfy the linear system
\begin{equation}
  \label{qAqI}
  \dot{q}_A = - \xi q_A, \;\;\; \dot{q}_I = \xi q_A - \mu q_I.
\end{equation}
These equations follow directly from Eqs.~(\ref{A}) and (\ref{I})
  by omitting the infection term in Eq.~(\ref{A}). 
The solution of \eqref{qAqI} with initial conditions $q_A(0) = 1, q_I(0) = 0$ reads
\begin{equation}
  \label{q}
      q(\tau) = \frac{1}{\mu - \xi} \left( \mu e^{-\xi \tau} - \xi
        e^{-\mu \tau}\right),
    \end{equation}
and taking the derivative we obtain the distribution of infection times as
\begin{equation}
  \label{p}
      p(\tau) = -\frac{d q}{d\tau} = \frac{\mu \xi}{\mu - \xi}
      \left(e^{-\xi \tau} - e^{-\mu \tau} \right).
    \end{equation}
 In the special case $\xi = \mu$ these
   expressions reduce to 
    \begin{equation}
      q(\tau) = (1+\mu \tau) e^{-\mu \tau}, \;\;\;\; p(\tau) =
      \mu^2 \tau e^{-\mu \tau}.
    \end{equation}
  It is instructive to interpret the distribution (\ref{p}) stochastically. As noted by Leung et al. \cite{Leung2018}, compartment models
    formulated in terms of ordinary differential equations implicitly assume that the transitions between the different compartments
    occur according to Markov processes with exponentially distributed waiting times. For the two-step process $A \to I \to R$
    described by the linear system (\ref{qAqI}) this implies that the total infection time can be written as a sum
    $\tau = t_\xi + t_\mu$ of two exponentially distributed waiting times $t_\xi$ and $t_\mu$ with parameters $\xi$ and $\mu$,
    which also leads to the probability density
    (\ref{p}). In general, for models with multiple
      compartments of infected individuals, the distribution of
      infection times is given by a convolution of
      exponential functions \cite{Keeling2008,Vazquez2021}.}

    \subsection{Basic reproduction number and fraction of asymptomatic infections}
    \label{Sec:Parameters}

\jk{To set the stage, we briefly recall the approach of Fraser et
  al. \cite{Fraser2004}. Denoting by $\beta(\tau)$ the
  infection rate of an individual at time
  $\tau$ since they were infected, the basic reproducion number is
  given by
  \begin{equation}
    \label{FraserR0}
    R_0 = \int_0^\infty d \tau \, \beta(\tau),
  \end{equation}
  and an outbreak occurs if $R_0 > 1$. Infectious individuals are
  asymptomatic at the time of infection, and remain in this state up
  to time $\tau$ with probability $\sigma(\tau)$. Accordingly, the
  fraction of asymptomatic infections is
  \begin{equation}
    \label{FraserTheta}
  \theta = \frac{\int_0^\infty d\tau \, \sigma(\tau)
    \beta(\tau)}{\int_0^\infty d\tau \, \beta(\tau)} = R_0^{-1} \int_0^\infty d\tau \, \sigma(\tau)
  \beta(\tau).
\end{equation}
In the following we determine the functions $\beta(\tau)$ and
  $\sigma(\tau)$ for the model defined by the differential equations (\ref{S}-\ref{R}).}    
The infectiousness function is given by $\beta(\tau) = \lambda q(\tau)$, and correspondingly
the basic reproduction number
is
\begin{equation}
  \label{R0}
    R_0 = \int_0^\infty d\tau \, \lambda q(\tau) =  \lambda \mathbb{E}(\tau) = \lambda
    \left(\frac{1}{\xi} + \frac{1}{\mu} \right)
    = R_0^A + R_0^I,
\end{equation}
where $\mathbb{E}(\tau)$ is the expectation of the probability density (\ref{p}) of infection times, and 
$R_0^A = \frac{\lambda}{\xi}$ and $R_0^I = \frac{\lambda}{\mu}$ denote the contributions of asymptomatic
and symptomatic infections to $R_0$ \cite{Dobrovolny2020,Ferretti2020,Tian2021,Chisholm2018}.
The fraction of asymptomatic infections is therefore
\begin{equation}
  \label{theta}
  \theta = \frac{R_0^A}{R_0} =  \frac{\mu}{\mu + \xi}.
\end{equation}
Combining (\ref{R0}) and (\ref{theta}) we can write
\begin{equation}
  \label{R0theta}
  R_0 = \frac{R_0^I}{1-\theta},
\end{equation}
which shows how the presence of (potentially undetected) asymptomatic infections increases the basic reproduction number
beyond the value $R_0^I$ estimated from symptomatic infections only.  

The second defining function of the theory of \cite{Fraser2004} is the probability for an infected individual
to still be asymptomatic after time $\tau$, which
can be obtained as
\begin{equation}
  \label{sigma}
  \sigma(\tau) =
  \frac{q_A(\tau)}{q(\tau)} = \frac{\mu - \xi}{\mu - \xi e^{-(\mu - \xi) \tau}}.
   \end{equation}
Interestingly, the function (\ref{sigma}) displays a qualitative change of behaviour when $\mu = \xi$ or $\theta = \frac{1}{2}$.
For $\theta < \frac{1}{2}$ the function approaches zero exponentially for large $\tau$, whereas for $\theta > \frac{1}{2}$  it attains
a nonzero limiting value of $2-\frac{1}{\theta} > 0$. Such a scenario was also discussed in \cite{Fraser2004}. Mathematically,
it reflects the fact that, when $\mu > \xi$, for exceptionally large values of the total infection time $\tau = t_\xi + t_\mu$
it is likely that $t_\xi \gg t_\mu$. In the degenerate case $\xi =
\mu$, $\sigma(\tau)$ decays algebraically as $(1+\mu \tau)^{-1}$.
It is readily checked that inserting (\ref{sigma}) into (\ref{FraserTheta})
  reproduces the relation (\ref{theta}) for the fraction of asymptomatic infections.

Together the relations (\ref{R0}) and (\ref{theta}) specify the ratios of the three rates $(\lambda, \xi, \mu)$ defining
the SAIR model. This implies that, up to an overall rescaling of time, the (linear and nonlinear)
behavior of the dynamical system (\ref{S}-\ref{R})
is fully determined by the two dimensionless epidemiological
parameters $R_0$ and $\theta$.

 \subsection{Asymptomatic recovery}
 \label{Sec:AtoR}

 In this section we assume that asymptomatically infected individuals
 recover at rate $\mu_A > 0$
 without previously developing symptoms. This changes Eqs.~(\ref{A})
 and (\ref{R}) into
 \begin{eqnarray}
   \dot{A} =  \frac{\lambda}{N} (A+I) S - (\xi + \mu_A) A \label{Arecovery}\\
   \dot{R}  =  \mu I + \mu_A A \label{Rrecovery}.
 \end{eqnarray}
The new rate $\mu_A$ can be expressed in terms of an additional
dimensionless parameter $\vartheta$ defined as the fraction of
individuals that never develop any symptoms. Since the
development of symptoms and recovery are competing processes
that both contribute to the loss of asymptomatic infected individuals, this
fraction is given by the ratio
\begin{equation}
  \label{vartheta}
  \vartheta = \frac{\mu_A}{\xi + \mu_A}.
\end{equation}
Importantly, as we now show, this is generally distinct from the
fraction $\theta$ of asymptomatic infections.

The computation of the epidemiological functions $q(\tau)$ and
  $\sigma(\tau)$ and of the parameters $R_0$ and $\theta$ proceeds
  along the lines of
  Sects.~\ref{Sec:Linear} and \ref{Sec:Parameters}. Replacing the
  first equation in (\ref{qAqI}) by $\dot{q}_A = -(\xi + \mu_A) q_A$,
  the probability $q(\tau)$ that an individual is still infectious at time
  $\tau$ becomes
  \begin{equation}
    \label{qRec}
    q(\tau) = \frac{1}{\mu - \mu_A - \xi} \left[(\mu - \mu_A) e^{-(\xi
        + \mu_A) \tau} - \xi e^{-\mu \tau} \right].
  \end{equation}
  Integrating this expression over $\tau$ we obtain
  \begin{equation}
    \label{R0Rec}
    R_0 = \frac{\lambda}{\mu} \left(\frac{\xi + \mu}{\xi + \mu_A}\right) = \lambda
    \left[ \frac{1}{\xi + \mu_A} + (1- \vartheta) \frac{1}{\mu}
    \right] = R_0^A + R_0^I.
  \end{equation}
  Similar to (\ref{R0}), the second equality expresses $R_0$ in terms of the expected
  infection times in the asymptomatic and the symptomatic compartment, which are given by $1/(\xi + \mu_A)$ and
  $1/\mu$, respectively. Since not all individuals develop
  symptoms, the latter contribution is weighted by a factor
  $1-\vartheta$. As in Sect.~\ref{Sec:Parameters}, the fraction of
  asymptomatic infections is then obtained as
  \begin{equation}
    \label{thetaRec}
    \theta = \frac{R_0^A}{R_0} = \frac{(\xi+\mu_A)^{-1}}{(\xi +
      \mu_A)^{-1} + (1 - \vartheta) \mu^{-1}} = \frac{\mu}{\mu+\xi},
    \end{equation}
 which, remarkably, does not depend on $\mu_A$. Based on Eqs.(\ref{vartheta})
 and (\ref{thetaRec}), we see that $\theta > \vartheta$ for $\mu > \mu_A$ and
 $\theta < \vartheta$ for $\mu < \mu_A$. Finally, the function
 $\sigma(\tau)$ is given by the expression
 \begin{equation}
   \label{sigmaRec}
 \sigma(\tau) = \frac{q_A(\tau)}{q(\tau)} = \frac{\mu - \mu_A -
   \xi}{\mu - \mu_A - \xi e^{-(\mu - \mu_A - \xi)\tau}},
\end{equation}
which approaches a positive limiting value for $\tau \to \infty$ when
$\mu > \mu_A + \xi$.

The model simplifies
considerably when symptomatic and asymptomatic individuals recover at
the same rate. In that case the functions $q(\tau)$ and
$\sigma(\tau)$ become pure exponentials, $\theta = \vartheta$, and $R_0 = \lambda/\mu$
independent of $\xi$. In fact, as pointed out in \cite{Ansumali2020},
for $\mu = \mu_A$ the total number of infected individuals $A + I$
satisfies standard SIR dynamics, and the transfer between the two
compartments has no consequences for the devlopment of the
epidemic. We will return to this case below in Sect.~\ref{Sec:Peak}.

\section{Peak of the epidemic}
\label{Sec:Peak}


In the following we focus on the peak number of symptomatic infections $I_\mathrm{max}$ as a measure for the severity
of an outbreak. In this section we derive a rigorous upper bound on
$I_\mathrm{max}$. \jk{Upper bounds are of particular significance for
  the evaluation of worst case scenarios.}

\jk{The key step in the analysis is the identification
  of a nontrivial conservation law for the dynamical system defined by
  Eqs.(\ref{S},\ref{A},\ref{I},\ref{R}). For this purpose we
  introduce the auxiliary quantity
\begin{equation}
  \label{Jdef}
  J(t) = A(t) + \phi I(t).
\end{equation}
Using (\ref{A}) and (\ref{I}), we see that with the choice 
\begin{equation}
  \phi = \frac{\xi}{\xi +
      \mu} = 1 - \theta
\end{equation}
$J(t)$} satisfies
\begin{equation}
  \label{J}
  \dot{J} = \dot{A} + \phi \dot{I} = \frac{\lambda}{N} (S - S^\ast) (A + I)
\end{equation}
where $S^\ast = \frac{N}{R_0}$. Thus $J(t)$ attains its peak value $J_\mathrm{max}$
at a time $t_J$ defined by $S(t_J) = S^\ast$. Dividing Eq.~(\ref{J}) by
Eq.~(\ref{S}) we obtain $\frac{dJ}{dS} = \frac{S^\ast}{S} -1$, which can be integrated to yield
$J - J_0 = S^\ast \ln \left( \frac{S^\ast}{S_0} \right) + S_0 - S$. Rearranging terms we conclude that the quantity
\begin{equation}
  \label{H}
  H \equiv J + S - S^\ast \ln S = A + S + (1-\theta) I - \frac{N}{R_0} \ln S
\end{equation}
is conserved under the dynamics.

At the time $t_I$ at which $I(t)$ attains its maximum value $I_\mathrm{max}$ we have $\dot{I} = 0$ and therefore
\begin{equation}
  \label{AIpeak}
 \jk{A(t_I)  = \frac{\mu}{\xi} I(t_I),}
\end{equation}
  which implies that
\begin{equation}
  \label{HtI}
  H(t_I) = \psi I_\mathrm{max} + \tilde{S} - S^\ast \ln \tilde{S}.
\end{equation}
Here $\tilde{S} = S(t_I)$ and
\begin{equation}
\psi =
\phi + \frac{\mu}{\xi} = 1 - \theta + \frac{\theta}{1 - \theta} \geq 1.
\end{equation}
Equating (\ref{HtI}) to the initial value $H(0) = N - S^\ast \ln N$ of $H$ corresponding to a completely susceptible population,
we conclude that
\begin{equation}
  \label{Imax}
  I_\textrm{max} = \frac{1}{\psi} \left[N - \tilde{S} + S^\ast \ln \left(\frac{\tilde{S}}{N} \right) \right].
\end{equation}
Observing finally that the quantity in square brackets on the right hand side of (\ref{Imax}) is maximized by
$\tilde{S} = S^\ast$, we arrive at the bound
\begin{equation}
  \label{bound}
  I_\textrm{max} \leq \frac{1}{\psi} J_\textrm{max} = \frac{N}{\psi} \left( 1 - \frac{1}{R_0} (1+\ln R_0) \right).
\end{equation}

\begin{figure}[t]
\begin{center}
\includegraphics[width=0.5\textwidth]{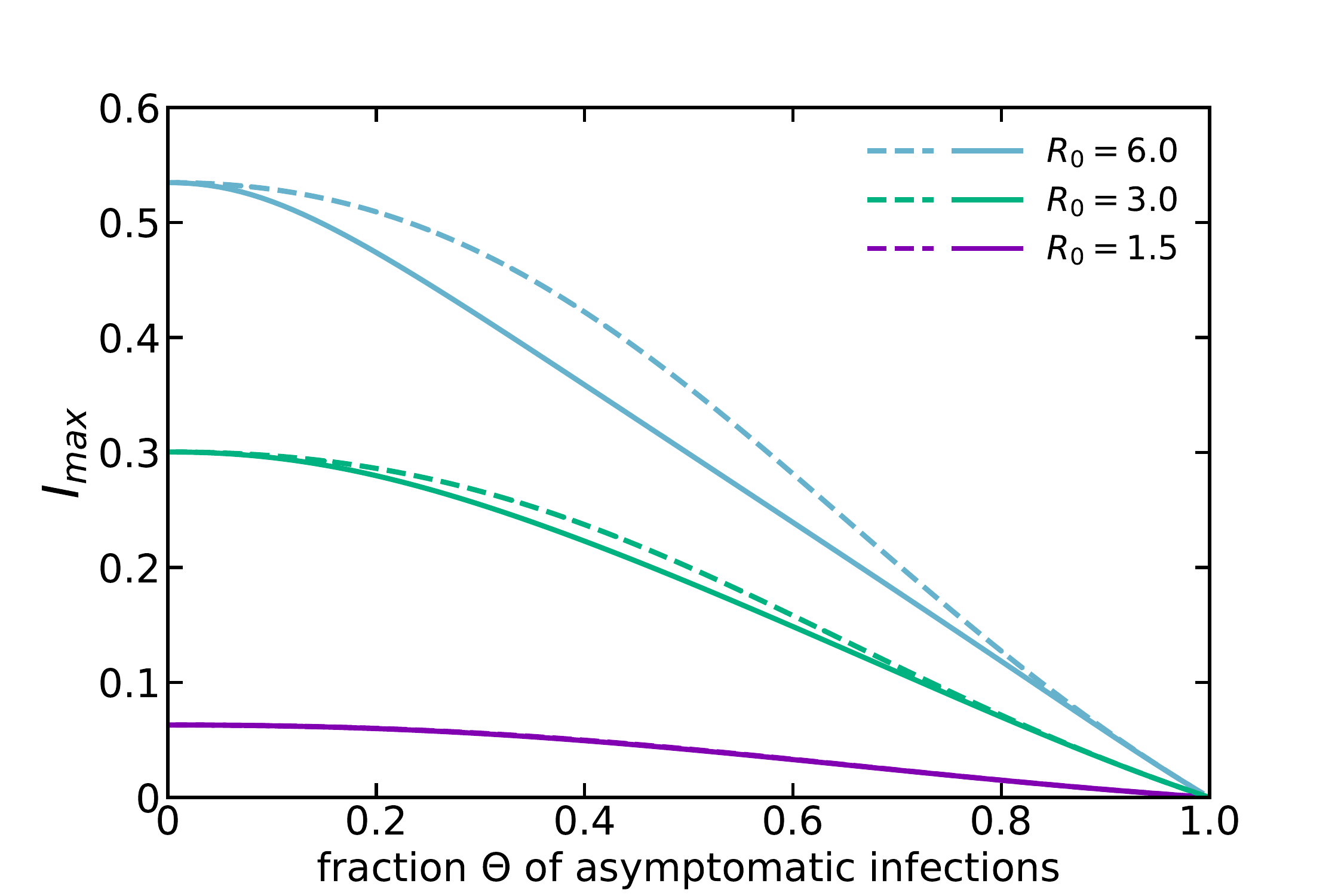}\includegraphics[width=0.5\textwidth]{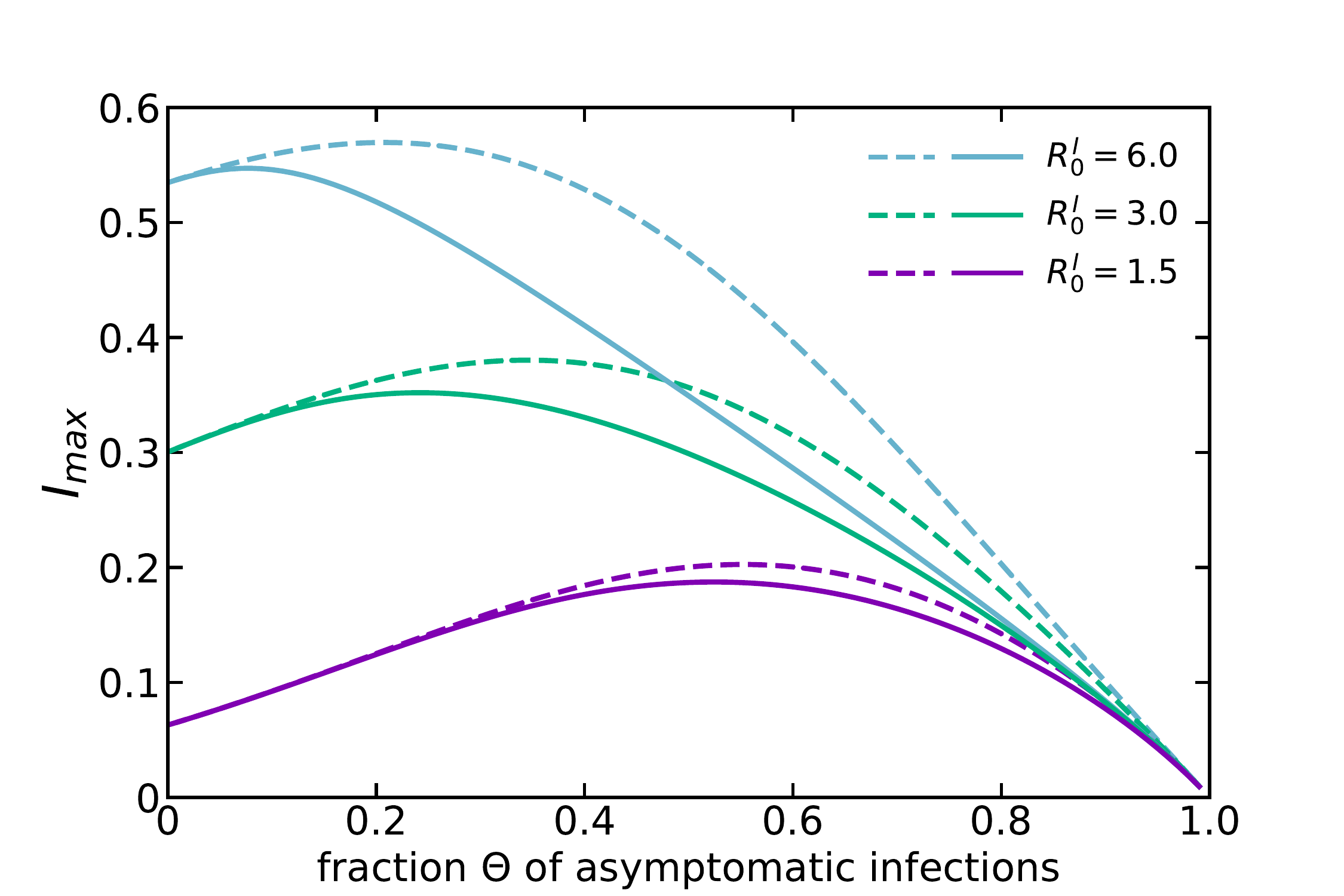}
\end{center}
\caption{\label{Fig1} The peak number of symptomatic cases $I_\mathrm{max}$
  as a function of the fraction $\theta$ of asymptomatic infections. Dashed
  lines show the analytic bound (\ref{bound}) and full lines show the results of the numerical solution of the
  dynamical system. In the left panel the total basic reproduction number $R_0$ is kept constant at the values
  $R_0 = 6$, 3 and 1.5 (from top to bottom). The right panel shows
  the corresponding behavior at constant $R_0^I = 6$, 3 and 1.5. In the latter case the total basic reproduction number
increases with the fraction of asymptomatic infections according to Eq.~(\ref{R0theta}).}
\end{figure}

Figure \ref{Fig1} illustrates the dependence of the bound (\ref{bound}) on the fraction of asymptomatic infections for two
scenarios, where either the total basic reproduction number $R_0$,
or the reproduction number due to symptomatic infections, $R_0^I$, is kept constant. In the first case $I_\mathrm{max}$
decreases monotonically with $\theta$, but the decrease is less than simply by a factor of $1-\theta$, as one might have
naively expected. In particular, for small $\theta$ the decrease is quadratic rather than linear in $\theta$.
Under the condition of constant $R_0^I$ the total basic reproduction number increases with increasing $\theta$ and diverges
for $\theta \to 1$, see Eq.~(\ref{R0theta}). In this case the bound displays a maximum at an intermediate value
of $\theta$ but nevertheless tends to zero for $\theta \to 1$.
The figure also shows exact results for $I_\mathrm{max}$ obtained by numerically integrating the system (\ref{S}-\ref{R}). The
analytic bound is seen to predict the qualitative behavior of $I_\mathrm{max}$ very well, but there are significant quantitative
deviations for large $R_0$ and intermediate values of $\theta$.

\jk{A similar bound can be derived for the model with asymptomatic
  recovery considered in Sect.~\ref{Sec:AtoR}. For convenience we
  restrict ourselves to the special case of equal recovery rates for
  symptomatic and asymptomatic individuals, $\mu = \mu_A$. Then the
  total number of infected individuals $A + I$ follows standard
  SIR dynamics, and its exact peak value $(A+I)_\textrm{max}$ is known
  \cite{Murray}. Using again the relation (\ref{AIpeak}) and the fact
  that
  \begin{equation}
    \label{A+I}
    A(t_I) + I(t_I) \leq (A+I)_\textrm{max} = N \left( 1 -
      \frac{1}{R_0} (1+\ln R_0) \right),
\end{equation}
we arrive at the bound
\begin{equation}
  \label{boundRec}
  I_\textrm{max} \leq (1-\theta) N \left( 1 - \frac{1}{R_0} (1+\ln R_0) \right).
\end{equation}
Since $1/\psi \geq 1 - \theta$ for $\theta \in [0,1]$, for a given
value of $R_0$ the right hand side of (\ref{boundRec}) is always smaller than
that of (\ref{bound}).}


\section{SAIR model with mitigation}
\label{Sec:Interventions}

\subsection{Dynamical equations with isolation and testing}

\begin{figure}[htb]
\begin{center}
\includegraphics[width=0.6\textwidth]{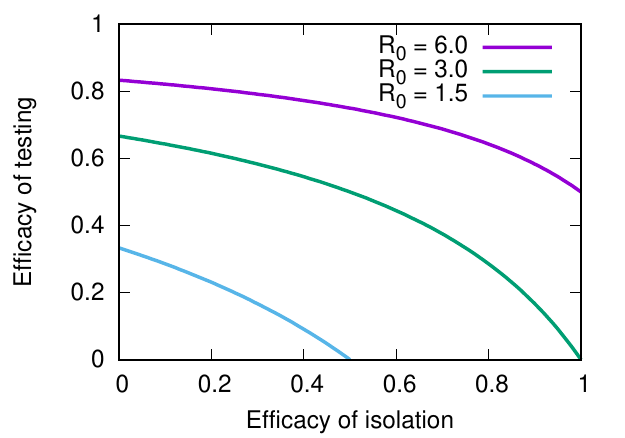}
\end{center}
\caption{\label{Fig2} The efficacy of testing $\epsilon_T$ required to reach the condition $\tilde{R}_0 = 1$ of
  controlling the outbreak is shown
  as a function of the efficacy of isolation $\epsilon_I$. The fraction of asymptomatic infections is
  $\theta = 1/3$, and the total basic reproduction number is $R_0=6$, 3 and 1.5 (from top to bottom). Points above the respective
  lines correspond to successful mitigation that is sufficient to control the outbreak.}
\end{figure}

The isolation of symptomatic cases and the testing of asymptomatic individuals are among the most important strategies
for containing an epidemic by non-pharmaceutical means \cite{Ferretti2020,Contreras2021,Tian2021,Fraser2004,Mazzitello2021,Mukhamadiarov2022}.
Adding the corresponding processes to the dynamical system (\ref{S}-\ref{R}) with isolation and testing rates
$\nu_I$ and $\nu_T$ leads to the extended model
\begin{eqnarray}
  \dot{S} =  - \frac{\lambda}{N} (A+I) S \label{SSS} \\
  \dot{A}  =  \frac{\lambda}{N} (A+I) S - (\xi + \nu_T) A \label{AAA} \\
  \dot{I}  =  \xi A - (\mu + \nu_I) I \label{III} \\
  \dot{T} = \nu_T A - \xi T \\
  \dot{Q} = \nu_I I + \xi T - \mu Q \label{Q} \\
  \dot{R}  =  \mu (I+Q),
\end{eqnarray}  
where the added compartments comprise asymptomatic cases that have been tested positively ($T$) and isolated (quarantined)
symptomatic cases ($Q$), respectively. Positively tested individuals develop symptoms at rate $\xi$ and isolated individuals
are removed at rate $\mu$. Since testing competes with the development of symptoms and isolation competes with removal, the
efficacy of the two interventions can be quantified by the ratios
  \begin{equation}
    \epsilon_T = \frac{\nu_T}{\nu_T + \xi}, \;\;\;\; \epsilon_I =
    \frac{\nu_I}{\nu_I + \mu}.
  \end{equation}
  To be precise, $\epsilon_T$ is the fraction of asymptomatic cases that are detected before developing symptoms, and
  $\epsilon_I$ is the fraction of symptomatic cases that are isolated before being removed. Repeating the analysis
  of Sections \ref{Sec:Linear} and \ref{Sec:Parameters} for the model with mitigation we arrive at the expression
  \begin{equation}
    \label{R0tilde}
  \tilde{R}_0 = \lambda \frac{\xi + \mu + \nu_I}{(\xi +
      \nu_T)(\mu+\nu_I)} = (1-\epsilon_T)[1 - \epsilon_I (1-
    \theta)] R_0
  \end{equation}
  for the basic reproduction number in the presence of mitigation. For $\epsilon_T = 0$ this reproduces a result of
  \cite{Fraser2004}, who define an outbreak to be controlled
  by the intervention if $\tilde{R}_0 \leq 1$. Figure~\ref{Fig2} delineates the region of controlled outbreaks
  in the $(\epsilon_I,\epsilon_T)$-plane for $\theta = \frac{1}{3}$ and different values of $R_0$. 

\subsection{Peak number of symptomatic cases}
\label{Sec:SAIR-PI}

Since symptomatic individuals require medical care irrespective of whether they have been isolated or not, the relevant
measure for the severity of an outbreak in the presence of mitigation is the peak in the total
number of symptomatic cases, i.e. the maximum $M_\mathrm{max}$ of $M(t) = I(t) + Q(t)$. To derive an upper bound on this
quantity, we again make use of the auxiliary quantity $J(t)$ defined in (\ref{Jdef}), where the factor $\phi$ is now given
by
\begin{equation}
    \phi = \frac{\xi + \nu_T}{\xi +
      \mu + \nu_I} = \frac{1}{(1-\epsilon_T)\left(1+\frac{\theta}{(1-\theta)(1-\epsilon_I)}\right)}.
  \end{equation}
Analogous to Sect.~\ref{Sec:Peak} it can be shown that, for $\tilde{R}_0 > 1$,
$J(t)$ attains its maximum when the number of susceptible individuals
reaches $S^\ast = \frac{N}{\tilde{R}_0}$, and the maximum value $J_\mathrm{max}$ is given by the expression in
Eq.~(\ref{bound}) with $R_0$ replaced by $\tilde{R}_0$.

In the following we specialize to the case without testing ($\nu_T = 0$). Then $T = 0$, and Eqs.~(\ref{III},\ref{Q})
can be combined to $\dot{M} = \xi A - \mu M$. At the time $t_M$ when $M$ reaches its maximum value we have $\dot{M} = 0$
and therefore
\begin{eqnarray}
  \label{Mbound}
  M_\mathrm{max} = M(t_M) = \frac{\xi}{\mu} A(t_M) = \frac{1-\theta}{\theta} A(t_M) \leq \frac{1-\theta}{\theta} J(t_M) \leq
  \nonumber \\
  \leq \frac{1-\theta}{\theta} J_\mathrm{max} = \frac{1-\theta}{\theta} N \left( 1 - \frac{1}{\tilde{R}_0} (1+\ln \tilde{R}_0) \right),
\end{eqnarray}
with $\tilde{R}_0 = R_0 (1 - \epsilon_I (1-\theta))$ for $\epsilon_T = 0$.

\begin{figure}[htb]
\begin{center}
\includegraphics[width=0.7\textwidth]{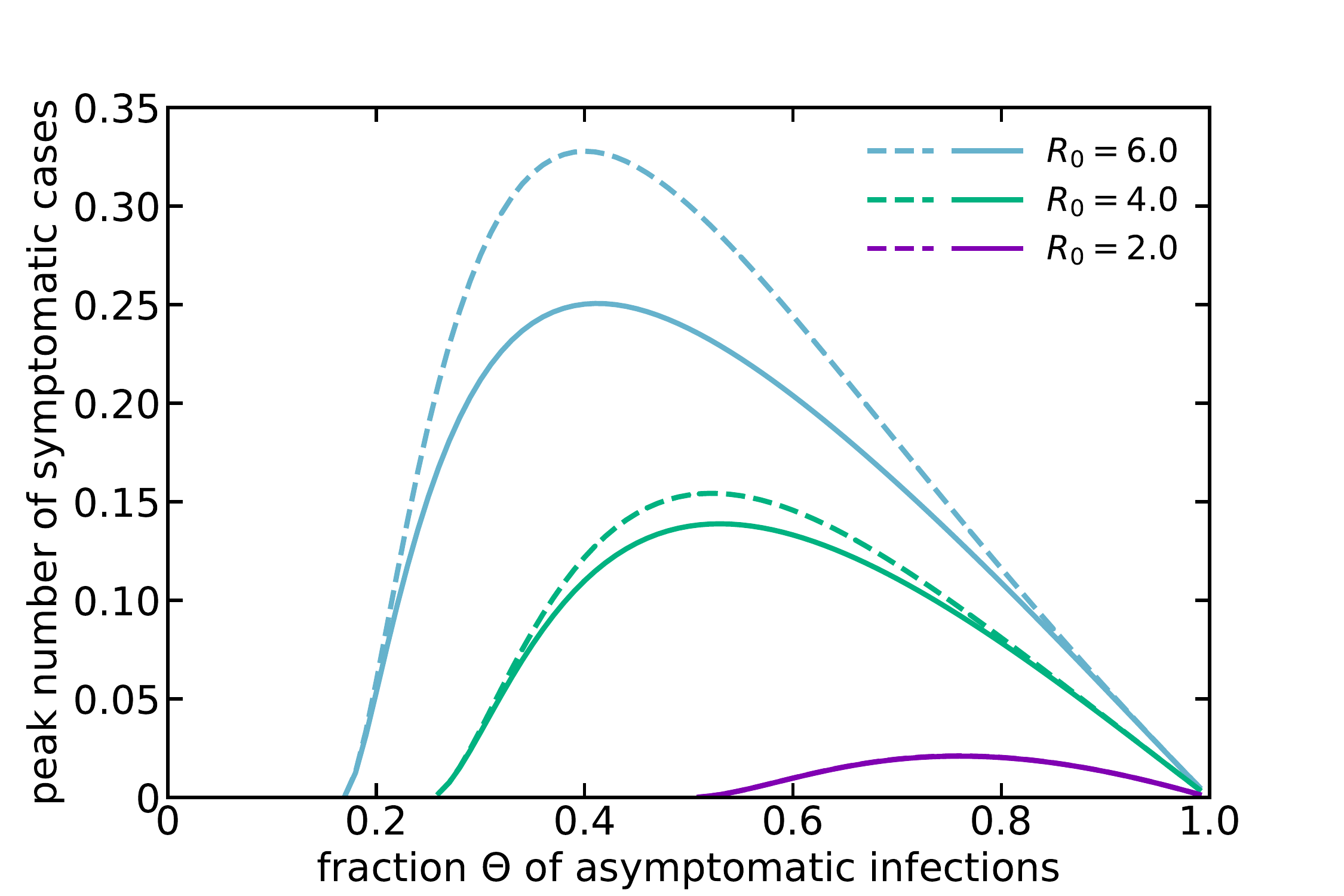}
\end{center}
\caption{\label{Fig3} Peak value of the number of symptomatic cases for the SAIR model with perfect isolation
  ($\epsilon_I = 1$) and no testing ($\epsilon_T = 0$). In this limit all symptomatic cases belong to the quarantine
  compartment. The effective basic reproduction number is $\tilde{R}_0 = \theta R_0$ and the outbreak is suppressed
  for $\theta <  \frac{1}{R_0}$. Dashed lines show the analytic bound (\ref{Mbound}) and full lines show the results of the
  numerical solution of the dynamical system.}
\end{figure}

Bounding $A$ by $J$ in (\ref{Mbound}) is clearly a crude approximation when $\phi$ is not small.
The bound is therefore expected to be most accurate when isolation is close to perfect, i.e. for $\epsilon_I \to 1$
or $\nu_I \to \infty$. In this limit symptomatically infected individuals are instantly transferred to the quarantine compartment,
which implies that the terms proportional to $I$ can be removed from Eqs.~(\ref{SSS},\ref{AAA}) and
Eq.~(\ref{III}) can be eliminated by replacing the term $\nu_I I$ in Eq.~(\ref{Q}) by $\xi A$. Thus
the limiting dynamics for perfect isolation and no testing reads 
\begin{eqnarray}
  \dot{S} =  - \frac{\lambda}{N} A S  \label{SPI} \\
  \dot{A}  =  \frac{\lambda}{N} A S - \xi A \label{API} \\
  \dot{Q} = \xi A - \mu Q \label{QPI} \\
  \dot{R}  =  \mu Q \label{RQI}.
\end{eqnarray}    
Figure \ref{Fig3} compares the bound (\ref{Mbound}) to the peak value $M_\mathrm{max} = Q_\mathrm{max}$ obtained by numerically
integrating the system (\ref{SPI}-\ref{RQI}). The agreement between the analytic and numerical results in this figure
is similar to Fig.~\ref{Fig1}. 

\section{Conclusions}

\label{Sec:Conclusions}

In this article we have introduced and studied a minimal compartment
model that allows to investigate the effects of asymptomatic
infections on the dynamics of an epidemic outbreak. The time evolution
of the model is fully specified by two dimensionless parameters, the
basic reproduction number $R_0$ and the fraction of asymptomatic
infections $\theta$. Empirical estimates of $\theta$ for SARS-CoV-2
vary widely \cite{Byambasuren2020,Alene2021}, but it is undisputed
that asymptomatic infections play a significant role in the current
pandemic \cite{Koelle2022}; a recent meta-analysis concluded that
$\theta \geq \frac{1}{3}$ \cite{Oran2021}.

Our approach is complementary to previous work based on renewal
theory, which determines the conditions for an outbreak from the
distributions of the times during which individuals are infectious
and/or symptomatic
\cite{Ferretti2020,Tian2021,Fraser2004,Grassly2008}. Whereas the
structure of our model constrains these distributions to
be of exponential form (Sect.~\ref{Sec:Linear}), it also enables us to go beyond
the initial stages of the outbreak and describe the full nonlinear
dynamics. Mitigation strategies such as testing and
isolation that differentiate between asymptomatic and symptomatic
cases can be incorporated in a straightforward
way \jk{(Sect.~\ref{Sec:Interventions}), and the
  effects of additional processes such as asymptomatic
  recovery can be evaluated systematically (Sect.~\ref{Sec:AtoR}).}
\jk{Despite the highly simplified and sketchy character of our
  mathematical framework \cite{Traulsen2022}, the explicit analytic expressions that we
  have derived may help to parametrize and interpret data-driven studies
  of more complex models with predictive capabilities \cite{Ferretti2020,Contreras2021,Tian2021}.}

A common property of epidemic compartment models that often makes
them analytically tractable is the existence of nontrivial
conservation laws \cite{Murray}, which is linked to an underlying Hamiltonian structure
\cite{Nutku1990,Ballesteros2020,Haas2022}. \jk{This feature provides a
  formal connection between mathematical epidemiology and theoretical physics.} Here we exploit such a conservation
law to derive rigorous analytic bounds on the peak number of
symptomatic infections $I_\mathrm{max}$, which serves as measure for the severity
of an outbreak and its societal consequences. Remarkably, under two
scenarios illustrated in Figs.~\ref{Fig1} and \ref{Fig3} we find that the peak
number of infections varies non-monotonically with $\theta$. This
reflects the dual role of asymptomatic cases: Although they
increase the severity of the outbreak by increasing the basic
reproduction number $R_0$, they do not contribute to the disease
burden. While it may seem obvious that $I_\mathrm{max} = 0$ for a fully
asymptomatic (`silent') epidemic ($\theta = 1$), the proof that this remains true even if $R_0$
diverges in the limit $\theta \to 1$ requires the explicit
mathematical analysis carried out in Sect.~\ref{Sec:Peak}. 

The SARS-CoV-2 virus has undergone significant evolution over the past two years, and the concomitant changes
in epidemiological parameters have contributed to the difficulty of controlling the pandemic \cite{Koelle2022}.
The observed increase of the basic reproduction number $R_0$ with each newly emerging variant was to be expected on
evolutionary grounds, but the selective forces acting on viral life history traits such as the time of the onset of symptoms
or the severity of the disease are complex and not well understood. Theoretical work addressing this question makes use
of SAIR-type models that are coupled to the evolutionary dynamics of the pathogen population \cite{Saad-Roy2020,Day2020}.
As such, a better analytic understanding of these models may contribute to forecasting the time course of future pandemics.

\appendix

  \section{SAIR models}
  \label{Sec:Related}

Epidemic compartment models with asymptomatic infections go back at
least to the work of Kemper in 1978 \cite{Kemper1978}. One may broadly
distinguish between models that allow for a loss of immunity, and
hence the approach to an endemic state, by
including a transfer from the removed to the susceptible compartment
\cite{Saad-Roy2020,Kemper1978,Robinson2013,Ottaviano2022}, and those
that assume irreversible immunization
\cite{Dobrovolny2020,Ansumali2020,Mukhamadiarov2022,Day2020,Debarre2007,Chladna2021}. To
position our contribution within the latter body of work, it is useful
to consider the following generalization of the model defined by Eqs.~(\ref{S}-\ref{R}):  
\begin{eqnarray}
\dot{S} =  - \frac{\lambda}{N} (\delta_A A+ \delta_I I) S  \label{SS} \\
  \dot{A}  =  (1-\eta) \frac{\lambda}{N} (\delta_A A+ \delta_I I) S - (\xi + \mu_A) A \label {AA}\\
  \dot{I}  =  \eta \frac{\lambda}{N} (\delta_A A+ \delta_I I) S + \xi A - \mu I \label{II} \\
  \dot{R}  =  \mu I + \mu_A A \label{RR}
\end{eqnarray}
Here the parameters $\delta_A, \delta_I \in [0,1]$ quantify the relative infectiousness of indidivuals in the compartments
$A$ and $I$, $\eta \in [0,1]$ is the probability that an infected individual develops symptoms immediately after infection, and
$\mu_A \geq 0$ is the rate at which $A$-individuals are removed without developing symptoms. The basic SAIR-model of
Sect.~\ref{Sec:SAIR} is recovered for $\delta_A = \delta_I = 1$ and $\eta = \mu_A = 0$. 

The dynamical system (\ref{SS}-\ref{RR}) includes several special cases that have been considered previously in the
literature. 

\begin{itemize}

\item For $\eta = \delta_A = \mu_A = 0$ the model reduces to the well-known SEIR-model, where the $A$-compartment
  (in this case denoted by $E$) contains \textit{exposed} individuals that are infected but not infectious \cite{Hethcote2000,Weinstein2020}.  

\item Ansumali et al. \cite{Ansumali2020} considered the model with
  $\eta = 0, \delta_A = \delta_I = 1, \mu_A = \mu$, \jk{which is a
    special case of the model with asymptomatic recovery considered in Sect.~\ref{Sec:AtoR}.}

\item Dobrovolny \cite{Dobrovolny2020} studied the model with $\xi = 0$ and used it to analyze data from the SARS-Cov-2 epidemics
  in several states of the USA.  

\item For $\delta_A = 1, \delta_I = \eta = \mu_A= 0$ the model reduces to the SAIR-model with perfect isolation of symptomatic individuals that
  is considered in Sect.~\ref{Sec:SAIR-PI}. In this case the $Q$-compartment of quarantined individuals plays the role
  of the $I$-compartment in Eqs.~(\ref{SS}-\ref{RR}). 

\end{itemize}

\end{document}